\documentclass[final]{siamltex}

\usepackage{amsfonts,amsmath}

\newtheorem{rem}{Remark}
\numberwithin{rem}{section}
\newenvironment{remark}{\begin{rem}\rm}{\end{rem}}

\newtheorem{corollary*}{Corollary.\ \ \kill}

\newtheorem{definition*}{Definition.\ \ \kill}

\title{On the spectrum of the Neumann problem for Laplace equation in
a domain with a narrow slit\thanks{This
        work was supported by RFBR grants nos. 99-01-01143 and
        99-01-00139.}
}

\author{Rustem R.~Gadyl'shin\thanks{Bashkir State Pedagogical
University, Ufa, Russia, ({\tt gadylshin@bspu.ru}).
}
        \and Arlen M.~Il'in\thanks{Institute of Mathematics and Mechanics
        of the Russian academy of Sciences, Ekaterinburg, Russia ({\tt iam@imm.uran.ru})}}

\begin{document}

\maketitle

\begin{abstract}
The Neumann problem in two-dimensional domain with a narrow slit
is studied. The width of the slit is a small parameter. The
complete asymptotic expansion for the eigenvalue of the
perturbed problem converging to a simple eigenvalue of the
limiting problem is constructed by means of the method of the
matched asymptotic expansions. It is shown that the regular
perturbation theory can formally be applied in a natural way up
to terms of order $\varepsilon^2$. However, the result obtained
in that way is false. The correct result can be obtained only by
means of inner asymptotic expansion.

Bibliography: 8 titles.

\end{abstract}



\pagestyle{myheadings} \thispagestyle{plain} \markboth{R.R.
Gadyl'shin AND A. M. Il'in}{On the spectrum of the Neumann
problem\ldots}

\section*{Introduction}
The Neumann problem in a two-dimensional domain with a narrow
slit is considered; it is called the perturbed problem in what
follows. The slit's width is a small parameter $\varepsilon$. In
the paper, we construct the complete asymptotics expansions for
an eigenvalue converging to a simple eigenvalue of the limiting
problem. The limiting problem is the Neumann problem in the
domain without the segment which the slit shrinks to. The
perturbed problem is singular; the asymptotics series in power
of the small parameter for the eigenfunction is valid everywhere
far from the endpoints of the segment and it fails near them.
Besides, the coefficients of the outer expansion have increasing
singularities near the segment's endpoints. Moreover, below (in
\S~3) we shall show that the regular perturbation theory can be
formally realized in a natural way up to a quantity
$\varepsilon^2$. However, it turns out that the results obtained
in such a way are not valid. Only using inner asymptotic
expansion allows us to get correct results. In this paper, the
construction of the asymptotics for an eigenvalue and of the
uniform asymptotics for an eigenfunction is carried out by the
method of matched asymptotics expansions \cite{Vd}-\cite{Ec}.

\section{Statement of the problem and formulation of the results}

Let $\Omega $ be a bounded simply connected domain in ${\mathbb{
R}}^2$ with infinitely differentiable boundary $\Gamma$, $\omega
_0$ be the interval $(0,1)$ in the axis $Ox_1$, $\overline
{\omega}_0\subset\Omega$, $\omega_\varepsilon=\{x:\, 0<x_1<1,\,
\varepsilon g_{-}(x_1)<x_2<\varepsilon g_{+}(x_1)\}$, where
$0<\varepsilon<<1$, $g_{\pm }\in C^\infty(\omega_0)$, $\pm
g_\pm>0$. We assume that in a neighbourhood of the endpoints of
the slit $\omega_\varepsilon$ its boundary lies on the
parabolas, i.e.,
$$
g_\pm(t) = \pm g^-t^{1/2}\quad\mbox{as $t<t_0$},\qquad g_\pm(t) =
\pm g^+(1-t)^{1/2}\quad\mbox{as $t>1-t_0$},\qquad g^\pm>0,
$$
where $t_0>0$ is some fixed number. We denote $\Omega_\delta
=\Omega\backslash\overline{\omega}_\delta$, $\delta \ge 0$,
$\gamma_\varepsilon=\partial\omega_\varepsilon$; $\gamma_0$ is
the cut $\{x:\, 0<x_1<1,\,x_2=0\}$ on the plain interpreted as
double-sided, $\Gamma_\delta=\Gamma\cup \gamma_\delta$. Under
the notation introduced the limiting and perturbed problems can
be written in the uniform way
\begin{equation}
-\Delta\phi_\delta = \lambda _\delta\phi_\delta, \quad
x\in\Omega_\delta, \qquad
\frac{\partial}{\partial\nu}\phi_\delta=0, \quad
x\in\Gamma_\delta, \label{1.1}
\end{equation}
where $\nu$ is the outer normal, $\delta=0$ corresponds to the
limiting problem, and $\delta =\varepsilon>0$ does to the
perturbed problem. It is convenient to consider the solutions of
both the perturbed and limiting problem in the class of
generalized solutions in Sobolev space $H_1$ (see, for instance,
\cite{Ld}). We use the notation $H_m(Q)$ for the Sobolev space
of functions on $Q$ whose derivatives of order less than or
equal to $m$ are square integrable.

Note that the Neumann boundary condition on the ``outer''
boundary $\Gamma$ are chosen for the sake of unambiguousness and
it is not principal for proofs used in paper. The only specific
(but not principal) consequence of this choice is that the
minimal perturbed eigenvalue equals zero.

We denote by $\Sigma_\delta$ the set of the eigenvalues of the
problem (\ref{1.1}). In the second section, we shall prove the
following statement.

\begin{theorem}\label{th1.1}
a) If $K$ is any compact set in the complex plane such that
$K\cap\Sigma_0=\emptyset$, then
$K\cap\Sigma_\varepsilon=\emptyset$ for all sufficiently small
$\varepsilon$;

b) If the multiplicity of $\lambda_0\in\Sigma_0$ equals $N$, then
$N$ eigenvalues of the perturbed problem (with multiplicities
taken into account) converge to $\lambda_0$.
\end{theorem}

We denote by  $S_-(t)$ and $S_+(t)$ the circles of radius $t$
and centers at the points $O_-=(0,0)$ and $O_+=(1,0)$,
respectively. For the sake of brevity we shall use the following
notations $x_-=x$, $x_+=((1-x_1),x_2)$, $(r_{\pm},\theta_{\pm})$
are associated polar coordinates. Below it will be shown that
the limiting eigenfunction $\phi_0$ normalized in
$H_0(\Omega_0)$ and associated with simple eigenvalue
$\lambda_0$ has the asymptotics (as $r_{\pm}\to0$)
$$
\phi_0(x) =\phi_0(O_\pm)+ d_\pm
r_\pm^{1/2}\cos\left(\frac{\theta_\pm}{2}\right)+O(r_\pm).
$$

The main statement of the paper reads as follows.
\begin{theorem}
The asymptotics for the eigenvalue $\lambda_\varepsilon$ of the
perturbed problem converging to a simple eigenvalue $\lambda_0$ of
the limiting problem and the asymptotics for the associated
eigenfunction have the form
\begin{align}
\allowdisplaybreaks \lambda_\varepsilon&=\sum_{j=0}^\infty
\varepsilon^j\lambda_j, \label{1.2}\\
\phi_\varepsilon(x)&=\sum^\infty_{j=0}\varepsilon^j\phi_j(x),
\qquad x\in\Omega_\varepsilon\backslash\left(S_+(\varepsilon)\cup
S_-(\varepsilon)\right),\label{1.3}\\
\phi_\varepsilon(x)&=\sum_{j=0}^\infty \varepsilon^jv_j^\pm
\left(\frac{x_\pm}{\left(g^\pm\varepsilon\right)^2}\right),\qquad
x\in S_\pm(2\varepsilon), \label{1.4}
\end{align}
\begin{equation}
\begin{aligned}
\lambda_1=&\lambda_0\int\limits_0^1 \left(
g_+(x_1)\phi_0^2(x_1,+0)-g_-(x_1)\phi_0^2(x_1,-0)\right)\,dx_1
\\
&- \int\limits_0^1 \left(g_+(x_1)\left(\frac{d}{
dx_1}\phi_0(x_1,+0)\right)^2-g_-(x_1)\left(\frac{d}{
dx_1}\phi_0(x_1,-0)\right)^2\right)\,dx_1,
\end{aligned}\label{1.5}
\end{equation}
\begin{equation}
\lambda_2=\frac{\pi}{8}\left(\left(d_{+}g^+\right)^2 +
\left(d_{-}g^-\right)^2\right)+\widetilde \lambda, \label{1.6}
\end{equation}
\begin{equation}
v_0^\pm(\xi)\equiv\phi_0(O_\pm),\qquad v_1^\pm(\xi)= d_{\pm}
g^{\pm}\mathrm{Re}\,\left(\xi_1+\mathrm{i}\xi_2-\frac{1}{4}\right)^{1/2}+
\phi_1(O_\pm), \label{1.7}
\end{equation}
\begin{equation}
\phi_2(x)=\frac{1}{8}\left(d_{+}\left(g^+\right)^2\psi_{+}(x)+
d_{-}\left(g^-\right)^2\psi_{-}(x)\right)+\widetilde\phi(x),\label{1.8}
\end{equation}
where  $\phi_1$, $\widetilde\phi$, and $\psi_{\pm}$ are the
functions satisfying the statements of Lemmas~\ref{lm3.2},
\ref{lm3.3}, and \ref{lm4.1}, respectively, $\widetilde\lambda$
is the constant determined by the equality (\ref{3.10}),
$\xi=(\xi_1,\xi_2)$, and $\mathrm{i}$ is the imaginary
unit.\label{th1.2}
\end{theorem}

The sections 3--8 are devoted to the construction and
justification of the asymptotics (\ref{1.2})--(\ref{1.8}) (i.e.,
to the complete proof of the Theorem~\ref{th1.2}). In the third
section, the coefficients  $\lambda_1$ and  $\phi_1$ are defined
by the regular theory of perturbation.  In the fourth section,
on the basis of the method of matched asymptotics expansions the
coefficients $\lambda_2$, $\phi_2$, and two first couples of the
coefficients $v^\pm_0$, $v^\pm_1$ for the inner expansion in the
neighbourhood of the slit's endpoints are determined. In the
fifth and sixth sections, we construct the complete outer and
inner expansions (\ref{1.4}) and  (\ref{1.3}), respectively. In
the seventh section, it is shown that they can be matched. In
the eighth section, the formally constructed asymptotics are
justified what completes the proofs of Theorem~\ref{1.2}. In the
concluding ninth, section we discuss the cases of other boundary
conditions on the boundary of the slit.

\section{Proof of the Theorem~\ref{th1.1}}
For a set $Q$ we denote by $(\bullet,\bullet)_Q$ the scalar
product in $H_0(Q)$. The solution of the boundary value problem
\begin{equation}
-\Delta u_\delta = \lambda u_\delta + f_\delta,\quad
x\in\Omega_\delta, \qquad
\frac{\partial}{\partial\nu}\phi_\delta=0, \quad
x\in\Gamma_\delta, \label{2.1}
\end{equation}
where $f_\delta\in H_0(\Omega_\delta)$ is the element $u_\delta\in
H_1(\Omega_\delta)$ satisfying the integral identity
\begin{equation}
(\nabla u_\delta,\nabla v)_{\Omega_\delta}= (\lambda
u_\delta+f_\delta,v)_{\Omega_\delta} \label{2.2}
\end{equation}
for each $v\in H_1(\Omega_\delta)$. Hereinafter, the function in
$H_0(\Omega_\varepsilon)$ are assumed to be continued by zero
inside $\omega_\varepsilon$ and they and the functions in
$H_0(\Omega_0)$ are identified with the elements of
$H_0(\Omega)$. By $\|\bullet\|_{m,Q}$ we denote the
$H_m(Q)$-norm.

Beforehand we prove an auxiliary lemma being a convenient variant
of well-known embedding theorems.

\begin{lemma}\label{lm2.1}
Let a function $w\in H_1(\Omega_\varepsilon)$, $\Pi(\alpha)$ be a
rectangle $\{x:\,-\alpha\le x_1\le 1+\alpha,\,\,|x_2|\le\alpha\}$,
$\Pi(\alpha,\varepsilon)=\Pi(\alpha)\cap\Omega_\varepsilon$,
$Q(\alpha)=\Pi(2\alpha)\backslash\Pi(\alpha)$. If parameters
 $\alpha$ and $\varepsilon$ are such that
$\Pi(2\alpha)\subset\Omega$ and
$\omega_\varepsilon\subset\Pi_\alpha$, then for all sufficiently
small $\varepsilon
>0$ the estimate
\begin{equation}
\| w\|^2_{0,\Pi(\alpha,\varepsilon)}\le2\| w\|^2_{0,Q(\alpha)}+
4\alpha^2\| w\|^2_{1,\Omega_\varepsilon}\label{2.3}
\end{equation}
holds.
\end{lemma}
\begin{proof}
At first let us consider the values $x_2\ge0$. We set
$\Pi_+(\alpha)=\Pi(\alpha)\cap\{x:\,x_2>0\},
\Pi_+(\alpha,\varepsilon)=\Pi(\alpha,\varepsilon)\cap\{x:\,x_2>0\},
Q_+(\alpha)=\Pi_+(2\alpha)\backslash \Pi_+(\alpha)$. Due to the
density of embedding $C^\infty(\overline{\Omega}_\varepsilon)$
in $H_1(\Omega_\varepsilon)$ we may suppose that $w\in
C^\infty(\overline{\Omega}_\varepsilon)$. Then for each $x_1\in
[-\alpha,1+\alpha]$ there exists a point $z\in [\alpha,2\alpha]$
such that
$$
|w(x_1,z)|\le\frac{1}{\alpha}\int\limits_\alpha^{2\alpha}
|w(x_1,\eta)|d\eta.
$$
Hence, for each point $(x_1,x_2)\in \Pi_+(\alpha,\varepsilon)$
$$
|w(x_1,x_2)|\le\frac{1}{\alpha}\int\limits_\alpha^{2\alpha}
|w(x_1,\eta)|d\eta + \left|\int\limits_{x_2}^{z}\frac{\partial
w}{\partial \eta}(x_1,\eta)d\eta\right|,
$$
\begin{equation}
|w(x_1,x_2)|^2\le\frac{2}{\alpha}\int\limits_\alpha^{2\alpha}
|w(x_1,\eta)|^2d\eta +4\alpha\int\limits_{x_2}^{z}|\nabla
w(x_1,\eta)|^2d\eta.\label{2.4}
\end{equation}
Let $\widetilde g_+(x_1)$ be the function that equals $g_+(x_1)$
for $0<x_1<1$ and vanishes for $-\alpha\le x_1\le0$ and for
$1\le x_1\le1+\alpha$. If we integrate inequality (\ref{2.4})
with respect to $x_2$ from $\varepsilon\widetilde g_+(x_1)$ to
$\alpha$, and after that we integrate the inequality obtained
with respect to $x_1$ from $-\alpha$ to $1+\alpha$, and we  take
into account that similar estimates are true for $x_2\le 0$,
then we get estimate (\ref{2.3}).
\end{proof}

\begin{lemma}\label{lm2.2} Let condition a) of
Theorem~\ref{th1.1} hold. Then

a) the statement a) of Theorem~\ref{th1.1} is valid;

b) for each  $\lambda\in K$,  $f_\varepsilon\in
H_0(\Omega_\varepsilon)$ the solutions of the boundary value
problem (\ref{2.1}) satisfy the uniform estimate
\begin{equation}
\|u_\varepsilon\|_{1,\Omega_\varepsilon}\,\le\, C \|
f_\varepsilon\|_{0,\Omega}. \label{2.5}
\end{equation}
\end{lemma}
\begin{proof}
It is easily seen that the equality (\ref{2.2}) yields the a
priori uniform estimate
\begin{equation}\label{2.6}
\| u_\varepsilon\|_{1,\Omega_\varepsilon}\,\le\, C_1 \left(\|
u_\varepsilon\|_{0,\Omega} + \| f_\varepsilon\|_{0,\Omega}\right).
\end{equation}
We prove the item b) by arguing by contradiction. Suppose that
there exist sequences $\lambda^{(n)}$ and $\varepsilon_n$ such
that the inequalities
\begin{equation}\label{2.7}
\| u_{\varepsilon_n}\|_{1,\Omega_\varepsilon}\,\ge\, n \|
f_{\varepsilon_n}\|_{0,\Omega}
\end{equation}
hold for $\lambda=\lambda^{(n)}$ and some $f_{\varepsilon_n}$.
Without loss of generality, we may assume that $\|
u_{\varepsilon_n}\|_{0,\Omega}=1$, $u_{\varepsilon_n}\to u_0$
weakly in $H_0(\Omega)$, $\varepsilon_n\to\varepsilon_0$ and
$\lambda^{(n)}\to\lambda^{(0)}\in K$ for $n\to\infty$. We also
assume that  $\varepsilon_0=0$. We shall prove both the items a)
and b) simultaneously. Indeed, if the item a) is wrong, then
there exist sequences of eigenfunctions $u_{\varepsilon_n}$ and
eigenvalues $\lambda^{(n)}\to\lambda^{(0)}\in K$ such that
$\varepsilon_n\to0$. Obviously, inequality (\ref{2.7}) is
correct for the eigenfunctions. If the item a) is valid,
$\varepsilon_0 \not=0$ and it is sufficiently small, then
$\lambda^{(0)}$ is not an eigenvalue of the problem for
$\varepsilon=\varepsilon_0$. Then the uniform estimate
(\ref{2.5}) follows from the well-known a priori estimates for
the solutions of the elliptic equations in a domain with a
smooth boundary. Thus, $\varepsilon_n\to 0$.

From (\ref{2.6}) and  (\ref{2.7}) it follows  that
\begin{equation}\label{2.8}
\| u_{\varepsilon_n}\|_{1,\Omega_{\varepsilon_n}} \,\le\, C_2.
\end{equation}
Observe that estimates (\ref{2.7}) and (\ref{2.8}) yield the
convergence to zero of $f_{\varepsilon_n}$ in
$H_0(\Omega)$-norm. In the proofs of this and next lemmas we
denote by $M$ any compact set $M\subset \overline{\Omega}$
separated from the segment $\omega_0$. We select the subsequence
from the sequence $u_{\varepsilon_n}$ which converges to the
function $u_0$ in $H_1$-norm on $M$. For this subsequence we use
the former notation $u_{\varepsilon_n}$. The existence of this
subsequence follows from the convergence to zero of
$f_{\varepsilon_n}$ in $H_0(\Omega)$-norm and from the
well-known a priori estimates for the solutions of an elliptic
equation in $M$. Note that $u_0\in H_1(\Omega_0)$ due to the
uniform boundedness (\ref{2.8}).

Let us show that $u_0$ is a solution of the limiting problem,
i.e., for each $v\in H_1(\Omega_0)$ the identity
\begin{equation}\label{2.9}
(\nabla u_0,\nabla v)_{\Omega_0}= (\lambda^{(0)}
u_0,v)_{\Omega_0}
\end{equation}
holds. By definition, the identity
\begin{equation}\label{2.10}
(\nabla u_{\varepsilon_n},\nabla v)_{\Omega_{\varepsilon_n}}=
(\lambda^{(n)}
u_{\varepsilon_n}+f_{\varepsilon_n},v)_{\Omega_{\varepsilon_n}}
\end{equation}
holds. By the weak convergence $u_{\varepsilon_n}\to u_0$ in
$H_0(\Omega)$ we deduce the convergence
\begin{equation}\label{2.11}
(\lambda^{(n)}
u_{\varepsilon_n}+f_{\varepsilon_n},v)_{\Omega_{\varepsilon_n}}\to
(\lambda^{(0)} u_0,v)_{\Omega_0}.
\end{equation}
For all $\mu>\varepsilon$ we have the obvious equality
\begin{equation}\label{2.12}
\begin{aligned}
(\nabla u_\varepsilon,\nabla v)_{\Omega_\varepsilon}-(\nabla
u_0,\nabla v)_{\Omega_0}=& (\nabla (u_\varepsilon-u_0),\nabla
v)_{\Omega_\mu}\\+ &  (\nabla (u_\varepsilon-u_0),\nabla
v)_{\Omega_\varepsilon\backslash\Omega_\mu} - (\nabla u_0,\nabla
v)_{\Omega_0\backslash\Omega_\varepsilon}.
\end{aligned}
\end{equation}
>From (\ref{2.12}), the arbitrariness in choosing  $\mu$, the
estimate (\ref{2.8}), and the convergence $u_{\varepsilon_n}$ in
$H_1$-norm on each compact set $M$ we derive the convergence
\begin{equation}\label{2.13}
(\nabla u_{\varepsilon_n},\nabla
v)_{\Omega_{\varepsilon_n}}\to(\nabla u_0,\nabla v)_{\Omega_0}.
\end{equation}
Assertions (\ref{2.10}), (\ref{2.11}), and (\ref{2.13}) yield
(\ref{2.9}).

Let us show that $u_0\not=0$. Suppose contrary, i.e., that
$u_{\varepsilon_n}\to0$ in $H_1$-norm on each compact set $M$.
Then from this assumption and estimate (\ref{2.3}), the
arbitrariness in choosing $\alpha$ and estimate (\ref{2.8}) we
deduce a convergence $u_{\varepsilon_n}\to0$ in $H_0(\Omega)$,
what contradicts the normalization of $u_{\varepsilon_n}$ in
$H_0(\Omega)$. Hence, $u_0$ is a nontrivial solution of the
limiting problem
$$
-\Delta u_0 = \lambda^{(0)}u_0, \quad x\in\Omega_0, \qquad
\frac{\partial}{\partial\nu}u_0=0, \quad x\in\Gamma_0,
$$
what is impossible since $\lambda^{(0)}\notin\Sigma_0$. The latter
contradiction proves the lemma.
\end{proof}

\begin{lemma}\label{lm2.3}
Let $K$ be the compact set described in the formulation of
Lemma~\ref{lm2.2} and $f_\varepsilon\to f_0$ in $H_0(\Omega)$ as
$\varepsilon\to0$. Then the solution of the perturbed problem
(\ref{2.1}) converges to the solution of the limiting problem as
$\varepsilon\to0$ in $H_1$-norm on each compact set  $M\subset
\overline{\Omega}$ separated from the segment $\omega_0$ and in
$H_0(\Omega)$-norm uniformly on $\lambda\in K$.
\end{lemma}
\begin{proof}
In view of estimate (\ref{2.5}) the proof of the convergence in
$H_1$-norm for each fixed $\lambda$ on each compact set $M$
reproduces the proof of previous lemma. From this convergence
and estimate (\ref{2.3}) we obtain the convergence in
$H_0(\Omega)$ for each fixed $\lambda$. In its turn, the latter
convergence and estimate (\ref{2.5}) imply uniform convergences
in required norms in obvious way.
\end{proof}

\emph{Proof of Theorem~\ref{th1.1}} Recall that the validity of
the item a) of Theorem~\ref{th1.1} was shown in
Lemma~\ref{lm2.2}. Thus, it remains to prove the item b). We
denote by $\Lambda(\mu)$ a closed circle in $\lambda$-plane with
radius $\mu$ and center at $\lambda_0$. We choose $\mu$
sufficiently small to satisfy $\Lambda(\mu)\cap
\Sigma_0=\{\lambda_0\}$. The existence of such $\mu$ follows
from the item a) of Theorem~\ref{th1.1}. Let $\phi_0$ be an
eigenfunction associated with $\lambda_0$. We set $f_0=\phi_0$,
$f_\varepsilon=f_0$ in $\Omega_\varepsilon$ and
$f_\varepsilon=0$ in $\omega_\varepsilon$. Then by
Lemma~\ref{lm2.3}
$$
\int\limits_{\partial\Lambda(\mu)}u_\varepsilon(\bullet,\lambda)
d\lambda\to\int\limits_{\partial\Lambda(\mu)}u_0(\bullet,\lambda)
d\lambda\not=0,\qquad\varepsilon\to0.
$$
This assertion by the arbitrariness in choosing $\mu$ yields the
existence of an eigenvalue $\lambda_\varepsilon$ of the perturbed
problem converging to $\lambda_0$.

Let us show that the total multiplicity of the eigenvalues of
the perturbed problem converging to $\lambda_0$ equals $N$. We
choose sequence $\varepsilon_n\to0$ such that for
$\varepsilon=\varepsilon_n$ there exist $L$ eigenfunctions
$\{\phi_\varepsilon^{(i)}\}_{i=1}^L$ associated with eigenvalues
converging to $\lambda_0$. There is no loss of generality in
assuming that they are orthonormalized  in $H_0(\Omega)$. By
analogy with the proofs Lemmas~\ref{lm2.2}, \ref{lm2.3} it is
easy to show the existence of the subsequence for which
$\phi_\varepsilon^{(j)}\to\phi_0^{(j)}\not=0$ in $H_0(\Omega)$,
where $\phi_0^{(j)}$ are orthonormalized in $H_0(\Omega)$
eigenfunctions of the limiting problem associated with
$\lambda_0$. Since the multiplicity of $\lambda_0$ equals $N$,
we have an inequality $L\le N$. Suppose that $L<N$. Then there
exists an eigenfunction $\phi_0^{(L+1)}$ of the limiting problem
orthogonal to $\phi_0^{(j)}$ for $j\le L$. We set
$$
f_\varepsilon=\phi_0^{(L+1)}-
\sum_{i=1}^L(\phi^{(L+1)}_0,\phi_\varepsilon^{(i)})_{\Omega}\phi_\varepsilon^{(i)},\quad
x\in\Omega_\varepsilon,\quad f_\varepsilon=0,\quad
x\in\omega_\varepsilon.
$$
By definition,
\begin{equation}\label{2.14}
(f_\varepsilon,\phi_\varepsilon^{(i)})_\Omega=0,\qquad i\le
L,\qquad f_\varepsilon\to f_0=\phi_0^{(L+1)},\quad
\varepsilon\to0.
\end{equation}
Lemma~\ref{lm2.3} and assertions (\ref{2.14}) yield
$$
0=\int\limits_{\partial\Lambda(\mu)}u_\varepsilon(\bullet,\lambda)d\lambda\to
\int\limits_{\partial\Lambda(\mu)}u_0(\bullet,\lambda)d\lambda\not=0,\qquad
\varepsilon\to0.
$$
Due to contradiction obtained, $L=N$. \quad\endproof

\begin{lemma}\label{lm2.4} Let $\lambda_0$ be a simple eigenvalue
of the limiting problem. Then an eigenfunction $\phi_\varepsilon$
associated with $\lambda_\varepsilon\to\lambda_0$ converges to an
eigenfunction $\phi_0$ of the limiting problem in $H_0(\Omega)$.
\end{lemma}
\begin{proof} Let $f_0=\phi_0$, $f_\varepsilon=f_0$ in
$\Omega_\varepsilon$ and $f_\varepsilon=0$ in
$\omega_\varepsilon$. Then by Lemma~\ref{lm2.3} and
Theorem~\ref{th1.1}, we get the convergence as $\varepsilon\to0$
$$
\alpha_\varepsilon\phi_\varepsilon=\int\limits_{\partial\Lambda(\mu)}u_\varepsilon(\bullet,\lambda)d\lambda\to
\int\limits_{\partial\Lambda(\mu)}u_0(\bullet,\lambda)d\lambda=\alpha_0\phi_0\not=0.
$$
Hence, $\phi_\varepsilon\to\phi_0$.
\end{proof}

In justification of the asymptotics for the eigenelements
constructed in the next sections we need uniform on $\lambda$
and $\varepsilon$ estimates for the solutions of the perturbed
problem for $\lambda$ close to a simple eigenvalue of the
limiting problem. By analogy with Lemmas~\ref{lm2.2},
\ref{lm2.3} one can prove
\begin{lemma}\label{lm2.5} Let $\lambda_0$ be a simple eigenvalue of
the limiting problem, $u_\varepsilon$ be the solution of
(\ref{2.1}) for $\lambda=\lambda_\varepsilon$, $\phi_\varepsilon$
be an associated eigenfunction and
$(u_\varepsilon,\phi_\varepsilon)_\Omega=0$. Then the estimate
$$
\| u_\varepsilon\|_{1,\Omega_\varepsilon}\,\le\, C \|
f_\varepsilon\|_{0,\Omega},
$$
holds, where the constant $C$ does not depend on $\varepsilon$.
\end{lemma}
\begin{lemma}\label{lm2.6} For $\lambda$ close to a simple eignevalue
$\lambda_0$ for the solution of the problem (\ref{2.1}) the
uniform estimates
\begin{align}
\|\lambda_\varepsilon-\lambda\|\,
|(u_\varepsilon,\phi_\varepsilon)_\Omega|&\le \|
f_\varepsilon\|_{0,\Omega}, \label{2.15}
\\
\| u_\varepsilon-(u_\varepsilon,\phi_\varepsilon)_\Omega
\phi_\varepsilon\|_{1,\Omega_\varepsilon} &\le C_1\left(\|
f_\varepsilon\|_{0,\Omega}+|\lambda-\lambda_\varepsilon|
\|u_\varepsilon\|_{0,\Omega}\right), \label{2.16}
\end{align}
where $\phi_\varepsilon$ is an eigenfunction of the perturbed
problem normalized in $H_0(\Omega)$.
\end{lemma}
\begin{proof}Substituting
$v=\phi_\varepsilon$ into assertion (\ref{2.2}), bearing in mind
the equality $(\nabla\phi_\varepsilon,\nabla
u_\varepsilon)_{\Omega_\varepsilon}=
\lambda_\varepsilon(\phi_\varepsilon,u_\varepsilon)_\Omega$ and
the fact that $\phi_\varepsilon$ and $\lambda_\varepsilon$ are
real-valued, we obtain estimate (\ref{2.15}). In its turn,
applying Lemma~\ref{lm2.5} to the function
$u_\varepsilon-(u_\varepsilon,\phi_\varepsilon)_\Omega\phi_\varepsilon$,
we obtain estimate (\ref{2.16}).
\end{proof}

\section{Regular perturbation theory}

Hereinafter, we deal with a simple eigenvalue $\lambda_0$, we
shall not fix this fact additionally in what follows.
Lemma~\ref{lm2.4} implies that the eigenfunction
$\phi_\varepsilon$ converges to $\phi_0$. For this reason, the
leading term of the asymptotics for $\phi_\varepsilon$ is
$\phi_0$. It is naturally to seek the asymptotics for
$\lambda_\varepsilon$ and $\phi_\varepsilon$ as (\ref{1.2}) and
(\ref{1.3}).

The boundary condition on $\gamma_\varepsilon$ in (\ref{1.1}) has
the form
\begin{equation}\label{3.1}
\varepsilon g'_\pm(x_1)\frac{\partial\phi_{\varepsilon}}{
\partial x_1}(x_1,\varepsilon g_\pm(x_1))-
\frac{\partial\phi_{\varepsilon}}{
\partial x_2}(x_1,\varepsilon g_\pm(x_1))=0.
\end{equation}

Substituting (\ref{1.2}) and (\ref{1.3}) into  (\ref{1.1}), due to
(\ref{3.1}) we arrive at the boundary value problems for the
coefficients $\phi_j$:

\begin{align}
& -(\Delta+\lambda_0)\phi_0 = 0,\quad x\in\Omega_0,\qquad
\frac{\partial\phi_0}{\partial \nu}=0, \qquad
x\in\Gamma_0,\label{3.2}
\\
&
\begin{aligned}
-(\Delta+\lambda_0)\phi_1 &= \lambda_1\,\phi_0,\quad
x\in\Omega_0,\qquad \frac{\partial\phi_1}{\partial \nu}=0,\quad
x\in\Gamma,
\\
\frac{\partial\phi_1}{\partial x_2}&=\left(g'_\pm\frac{\partial}{
\partial x_1}-g_\pm\frac{\partial^2}{\partial
x_2^2}\right)\phi_0, \qquad x_1\in\omega_0, \quad x_2=\pm0,
\end{aligned}\label{3.3}
\\
&\begin{aligned} -(\Delta+\lambda_0)\phi_2 &=
\lambda_1\,\phi_1+\lambda_2\phi_0,\quad x\in\Omega_0,\qquad
\frac{\partial\phi_2}{\partial \nu}=0,\quad x\in\Gamma,\\
\frac{\partial\phi_2}{\partial x_2}&=\left(g'_\pm\frac{\partial}{
\partial x_1}-g_\pm\frac{\partial^2}{\partial
x_2^2}\right)\phi_1\\ & +\left(g_\pm g'_\pm\frac{\partial^2}{
\partial x_1\partial x_2}-\frac{1}{
2}g^2_\pm\frac{\partial^3}{\partial x_2^3} \right)\phi_0, \qquad
x_1\in\omega_0, \quad x_2=\pm0.
\end{aligned}\label{3.4}
\end{align}

By definition, the eigenfunction $\phi_0$ is a solution of the
boundary value problems (\ref{3.2}). Let us show that there exist
functions $\phi_1,\, \phi_2\in H_1(\Omega_0)$, being solutions of
the boundary value problems (\ref{3.3}) and (\ref{3.4}) for some
constants $\lambda_1$, $\lambda_2$.

With $\Omega_0$ being a domain with a cut, hereinafter by
$\overline\Omega_0$ we mean the set
$\Omega_0\cup\gamma_0\cup\Gamma\cup\{O_-;O_+\}$. Let
$(r,\,\theta)$ be polar coordinates. We denote by
$\Psi_k(\theta)$ a linear combination of $\sin(j\theta/2)$ and
$\cos(j\theta/2)$, where $0\le j\le k$ and $j\equiv k\pmod 4$.
The asymptotics for the solution of equation (\ref{2.1}) near
the endpoints of the cut $\gamma_0$ was studied in \cite{Kn}.
Let us formulate the lemma concerning the asymptotics for the
solution which will be used below.

\begin{lemma}\label{lm3.1} Let functions $f\in H_1(\Omega_0)\cap
C^\infty(\overline\Omega_0\backslash\{O_-;O_+\})$ and $h_\pm\in
C^\infty(\omega_0)$ have asymptotics
\begin{align}
f(x)&=\sum_{k=0}^\infty r^{k/2}\Psi_k(\theta), \label{3.5}
\\
h_\pm(x_1)&=\sum_{k=-1}^\infty (\pm 1)^kb_k x_1^{k/2}, \label{3.6}
\end{align}
as $r\to0$ and $x_1\to0$ and similar expansions as
$(1-x_1)^2+x_2^2\to0$. Further, let $u\in H_1(\Omega_0)$ be a
solution of the boundary value problem
\begin{align*}
-(\Delta+\lambda)u &= f,\quad x\in\Omega_0,\qquad \frac{\partial
u}{\partial\nu}=0, \quad x\in\Gamma,
\\
\frac{\partial u}{\partial x_2}& =h_\pm,\quad x_1\in\omega_0,\quad
x_2=\pm0.
\end{align*}
Then $u\in C^{\infty}(\overline\Omega_0\backslash\{O_-;O_+\})$ has
the asymptotics of the form (\ref{3.5}) in the vicinity of the
endpoints of the cut $\gamma_0$.
\end{lemma}

Hereinafter, the asymptotics series are assumed to be infinitely
differentiable with respect to the variables $x_1$ and $x_2$.

\begin{corollary*}
The eigenfunction $\phi_0$ has the expansion of the form
(\ref{3.5}) in the vicinity of the endpoints of the cut
$\gamma_0$, where $\Psi_0=\phi_0(O_\mp)$,
$\Psi_1(\theta)=d_\mp\cos(\theta/2)$ in the vicinity of the left
and right endpoints.
\end{corollary*}

\begin{lemma}\label{lm3.2} There exists a function $\phi_1\in
H_1(\Omega_0)\cap
C^\infty(\overline\Omega_0\backslash\{O_-;O_+\})$, having
asymptotics (\ref{3.5}) that is a solution of the boundary value
problem (\ref{3.3}) for $\lambda_1$ determined by the equality
(\ref{1.5}) and is orthogonal to $\phi_0$ in $H_0(\Omega_0)$.
\end{lemma}
\begin{proof} It follows from Corollary to Lemma~\ref{lm3.1} that
the right hand side of equation (\ref{3.3}) satisfies all
assumptions of Lemma~\ref{lm3.1}. By the equation in
(\ref{3.2}),
\begin{equation}\label{3.7}
\left(g'_\pm\frac{\partial}{\partial x_1}-
g_\pm\frac{\partial^2}{\partial x_2^2}\right)\phi_0
\Big|_{x_2=\pm0}= \left(\frac{\partial}{\partial
x_1}\left(g_\pm\frac{\partial}{\partial
x_1}\right)+\lambda_0\right)\phi_0 \Big|_{x_2=\pm0}.
\end{equation}
From (\ref{3.7}) and Corollary to Lemma~\ref{lm3.1} we deduce
that the right hand side of the boundary condition in
(\ref{3.3}) satisfies the hypothesis of Lemma~\ref{lm3.1}.
Therefore, for each $\lambda$ distinct from an eigenvalue there
exists a solution of a boundary value problem
\begin{align*}
-(\Delta+\lambda)u &= 0,\quad x\in\Omega_0,\qquad \frac{\partial
u}{\partial \nu}=0,\quad x\in\Gamma,
\\
\frac{\partial u}{\partial x_2}&=\left(g'_\pm\frac{\partial}{
\partial x_1}-g_\pm\frac{\partial^2}{\partial x_2^2}\right)\phi_0,
\qquad x_1\in\omega_0, \quad x_2=\pm0,
\end{align*}
satisfying the statement of Lemma~\ref{lm3.1}. Representing the
solution of the problem (\ref{3.3}) in the form $\phi_1=u+w$, we
obtain the following boundary value problem for the function $w$
\begin{equation}\label{3.8}
-(\Delta+\lambda_0)w =(\lambda_0-\lambda)u+
\lambda_1\,\phi_0,\quad x\in\Omega_0, \qquad \frac{\partial
w}{\partial \nu}=0,\quad x\in\Gamma_0.
\end{equation}
The necessary and sufficient solvability condition for
(\ref{3.8}) is the orthogonality in $H_0(\Omega)$ of the right
hand side and $\phi_0$; we achieve it by a suitable choice of
the constant $\lambda_1$. Let us assume that $\lambda_1$ is
chosen exactly in this way. Since the right hand side of the
equation in (\ref{3.8}) belongs to $H_1(\Omega_0)\cap
C^\infty(\overline\Omega_0\backslash\{O_-;O_+\})$, it follows
that, by Lemma~\ref{lm3.1}, the function $\phi_1$ satisfies the
statements of the lemma being proved. Integrating by parts the
left hand side of the equality
$$
-\left((\Delta+\lambda_0)\phi_1,\phi_0\right)_{\Omega_0}=
\lambda_1(\phi_0,\phi_0)_\Omega=\lambda_1
$$
and taking into account (\ref{3.7}), we have
\begin{align*}
\lambda_1=\int\limits_0^1 &\left(\phi_0(x_1,+0) \left(\frac{d}{ d
x_1}\left(g_+(x_1)\frac{d}{d
x_1}\right)+\lambda_0\right)\phi_0(x_1,+0)\right.
\\
&-\left.\phi_0(x_1,-0) \left(\frac{d}{d
x_1}\left(g_-(x_1)\frac{d}{d
x_1}\right)+\lambda_0\right)\phi_0(x_1,-0)\right)\,dx_1.
\end{align*}
In its turn, integrating by parts the right hand side of latter
equality one can obtain relation (\ref{1.5}). Since the function
$\phi_1$ is defined up to the term $\alpha\phi_0$, for
definiteness we choose the constant $\alpha$ on the basis of the
orthogonality condition $(\phi_1,\phi_0)_\Omega=0$.
\end{proof}

Observe that by Corollary to Lemma~\ref{lm3.1} and
Lemma~\ref{lm3.2} the right hand side in equation (\ref{3.4})
satisfies all assumptions of Lemma~\ref{lm3.1}, and by
(\ref{3.2}) and (\ref{3.3})
\begin{equation}\label{3.9}
\begin{aligned}
\left(g'_\pm\frac{\partial}{
\partial x_1}-g_\pm\frac{\partial^2}{\partial
x_2^2}\right)\phi_1\Big|_{x_2=\pm0} &+\left(g_\pm
g'_\pm\frac{\partial^2}{
\partial x_1\partial x_2}-\frac{1}{
2}g^2_\pm\frac{\partial^3}{\partial x_2^3}
\right)\phi_0\Big|_{x_2=\pm0}
\\
&=\left(\frac{\partial}{\partial x_1}
\left(g_\pm\frac{\partial}{\partial
x_1}\right)+\lambda_0\right)\phi_1 \Big|_{x_2=\pm0}.
\end{aligned}
\end{equation}
Hence, by Lemma~\ref{lm3.2} the boundary condition in
(\ref{3.4}) satisfies the hypothesis of Lemma~\ref{lm3.1}.
Reproducing the proof of Lemma~\ref{lm3.2} and taking into
account the equality (\ref{3.9}), we get the validity of the
following analog of Lemma~\ref{lm3.2}.

\begin{lemma}\label{lm3.3} There exists a function
$\widetilde\phi\in H_1(\Omega_0)\cap
C^\infty(\overline\Omega_0\backslash\{O_-;O_+\})$, having
asymptotics (\ref{3.5}), and being a solution of the boundary
value problem (\ref{3.4}) for $\lambda_2=\widetilde \lambda$,
defined by the equality
\begin{equation}\label{3.10}
\begin{aligned}
\widetilde\lambda=&\lambda_1\int\limits_0^1 \left(
g_+(x_1)\phi_0^2(x_1,+0)-g_-(x_1)\phi_0^2(x_1,-0)\right)\,dx_1
\\
&+\lambda_0\int\limits_0^1 \left(
g_+(x_1)\phi_0(x_1,+0)\phi_1(x_1,+0)-
g_-(x_1)\phi_0(x_1,-0)\phi_0(x_1,-0)\right)\,dx_1
\\
&- \int\limits_0^1 \left(g_+(x_1)\frac{d}{
dx_1}\phi_0(x_1,+0)\frac{d}{ dx_1}\phi_1(x_1,+0)\right.
\\
&\quad\left.-g_-(x_1)\frac{d}{dx_1}\phi_0(x_1,-0)\frac{d}{
dx_1}\phi_1(x_1,-0)\right)\,dx_1.
\end{aligned}
\end{equation}
\end{lemma}
\begin{remark}\label{rm3.1} Below, the
construction (and justification) of complete asymptotics
expansions for the eigenelements will imply that the coefficients
$\phi_1$ and $\lambda_1$ obtained above are correct. From the
formal point of view, by Lemma~\ref{lm3.3} it can be set
$\phi_2=\widetilde\phi$ and $\lambda_2=\widetilde\lambda$.
However, as we shall show  below, the values
$\lambda_2=\widetilde\lambda$ and $\phi_2=\widetilde\phi$ are
wrong.
\end{remark}

\section{Construction of the second terms of the asymptotics
by the method of matched asymptotics expansions}

In order to define correct terms $\lambda_2$ and $\phi_2$ one
should use inner asymptotics expansions near the endpoints of
the segment. The form of this expansions is defined by already
constructed $\phi_0$ and $\phi_1$ (in accordance with the method
of matched asymptotics expansions). By Lemmas 3.1, 3.2, and
Corollary to Lemma 3.1 the asymptotics
\begin{equation}\label{4.1}
\phi_0(x) =\phi_0^\pm+ d_\pm
r_\pm^{1/2}\cos\left(\frac{\theta_\pm}{2}\right)+O(r_\pm),\qquad
\phi_1(x) = \phi_1^\pm+O\left(r_\pm^{1/2}\right),
\end{equation}
hold at the endpoints of the cut, where
$\phi_i^\pm=\phi_i(O_\pm)$.

In the vicinity of the endpoints of the slit,  we seek
asymptotics for the solution in the form of a power series in
$\varepsilon$ whose coefficients are functions depending on the
scaled (inner) variables
$\xi=x\left(g^-\right)^{-2}\varepsilon^{-2}$ in the vicinity of
the left endpoint and
$\xi_1=(1-x_1)\left(g^+\right)^{-2}\varepsilon^{-2}$,
$\xi_2=x_2\left(g^+\right)^{-2}\varepsilon^{-2}$ in the vicinity
of the right endpoint. Rewriting asymptotics for $\phi_j$ at the
ends of the slit in terms of the inner variables, we obtain that
\begin{equation}\label{4.2}
\phi_0(x)+\varepsilon\phi_1(x) =\phi_0^\pm+\varepsilon\left(
d_{\pm} g^{\pm}\rho^{1/2}\cos(\theta/2)-\phi_1^\pm\right)
+O(\varepsilon^2),
\end{equation}
where $(\rho,\theta)$ are polar coordinates in the plane
$\xi=(\xi_1,\xi_2)$.

Equality (\ref{4.2}) suggests that in the vicinity of the
endpoints of the slit the eigenfunction expressed in terms of
the inner variables must have the form
\begin{equation}\label{4.3}
\phi_\varepsilon(x)=v_0^\pm(\xi)+\varepsilon v_1^{\pm}(\xi) +
O(\varepsilon^2).
\end{equation}

The boundary value problems for $v_j^\pm$ are obtained in the
standard way \cite{Il1}, \cite{Il2}. We substitute (\ref{4.3})
and (\ref{1.2}) into (1.1) and pass to the inner variables in
the equation and boundary conditions bearing in mind that near
the ends the equation of the slit has the form $\xi_1=\xi_2^2$.
Equating the coefficients of the least powers of $\varepsilon$,
we get the following boundary value problems
\begin{equation}\label{4.4}
\Delta v_j^\pm = 0,\quad\xi\in\Pi, \qquad \frac{\partial
v_j^\pm}{\partial\nu} =0,\quad \xi\in\partial\Pi,
\end{equation}
where $\Pi=\{\xi:\,\xi_1<\xi_2^2\}$.

Due to relation (\ref{4.2}) (and the ideology of the method of
matched asymptotics expansions) the solutions of (\ref{4.4})
must have asymptotics
\begin{equation}\label{4.5}
v_0^{\pm}(\xi) = \phi_0^{\pm} +o(1),\qquad v_1^{\pm}(\xi) =
d_{\pm} g^{\pm}\rho^{1/2}\cos(\theta/2)+\phi_1^\pm +o(1), \quad
\rho\to\infty.
\end{equation}

It is easy to see that the functions (\ref{1.7}), where the cut
is made along the ray $(1/4,\infty)$ of the real axis, are the
solutions of the problem (\ref{4.4}), having asymptotics
(\ref{4.5}). Moreover, from (\ref{1.7}) it follows that
\begin{equation}\label{4.6}
v_1^{\pm}(\xi) = d_\pm
g^{\pm}\rho^{1/2}\cos(\theta/2)+\phi_1^\pm-\frac{1}{8}d_\pm
g^{\pm}\rho^{-1/2}\cos(\theta/2) + O(\rho^{-3/2}), \quad
\rho\to\infty.
\end{equation}

Rewriting now the asymptotics for $v_0^\pm+\varepsilon v_1^{\pm}$
at infinity in terms of the outer variables $x$, from (\ref{1.7})
and (\ref{4.6}) we deduce the leading terms of the asymptotics at
the endpoints of the slit for the coefficient $\phi_2$ of the
series (\ref{1.3})
\begin{equation}\label{4.7}
\phi_2(x)=-\frac{1}{}d_-
\left(g^-\right)^2r^{-1/2}\cos(\theta/2)+O(1),\quad r\to0
\end{equation}
at the left endpoint and a similar form holds an the right
endpoint.
\begin{remark}\label{rm4.1} From the asymptotics
(\ref{4.7}) it follows  that  $\phi_2$ does not belong to the
class $H_1(\Omega_0)$ (in the general case
$|d_{+}|+|d_{-}|\neq0$). For this very reason the formally
consistent regular second terms mentioned in Remark~\ref{rm3.1}
leads one to wrong values of the required quantities.
\end{remark}

Let us proceed to the construction of the correct terms $\phi_2$
and $\lambda_2$ of the expansions (\ref{1.2}) and (\ref{1.3}). To
this end, one should modify $\widetilde\phi$ and
$\widetilde\lambda$ constructed in Lemma~\ref{3.3} by taking into
account the asymptotics (\ref{4.7}). In other words, we should add
the singular term $\mathrm{const} r^{-1/2}\cos(\theta/2)$ to the
function $\widetilde\phi(x)$ near the left endpoint of the slit
and a similar term near the right endpoint of the slit. However,
this is not sufficient. In order that the function $\phi_2$
remains the solution of the problem (\ref{3.4}) we must add an
additional term belonging to $H_1(\Omega_0)$.

We use the notation $\chi(t)$ for the infinitely differentiable
cut-off function equal to one for $t<c$ and to zero for $t>2c$,
where $c<1/2$ is sufficiently small number so that the closed
circles of radius $2c$ with centers at $(0,0)$ and $(1,0)$ lie in
$\Omega$.
\begin{lemma}\label{lm4.1} There exist functions
\begin{equation}\label{4.8}
\psi_{\pm}(x)=\chi(r_{\pm})r_{\pm}^{-1/2}\cos(\theta_{\pm}/2)+
\widetilde\psi_\pm(x),
\end{equation}
where $\widetilde\psi_\pm\in H_1(\Omega_0)\cap
C^\infty(\overline\Omega_0\backslash\{O_-;O_+\})$, that are
solutions of a boundary value problem
\begin{equation}\label{4.9}
-(\Delta+\lambda_0)\psi_\pm=\lambda_{\pm}\phi_0, \quad
x\in\Omega_0,\qquad \frac{\partial\psi_\pm}{\partial\nu}=0,\quad
x\in\Gamma_0
\end{equation}
for
\begin{equation}\label{4.10}
\lambda_\pm=-\pi d_\pm
\end{equation}
\end{lemma}
\begin{proof} Let us seek $\widetilde\psi_\pm$ in the form
$$
\widetilde\psi_{\pm}(x)=-\frac{\lambda_0}{2}\chi(r_{\pm})r_{\pm}^{3/2}
\cos(\theta_{\pm}/2)+\widehat\psi_\pm(x).
$$
Substituting the expression (\ref{4.8}) into (\ref{4.9}), we
arrive at the following problem for $\widehat\psi_\pm$:
\begin{equation}\label{4.11}
-(\Delta+\lambda_0)\widehat\psi_\pm=\lambda_{\pm}\phi_0+f_\pm,
\quad x\in\Omega_0,\qquad
\frac{\partial\widehat\psi_\pm}{\partial\nu}=0,\quad x\in\Gamma_0,
\end{equation}
where $f_\pm\in H_1(\Omega_0)\cap
C^\infty(\overline\Omega_0\backslash\{O_-;O_+\})$ have the
asymptotics (\ref{3.5}). A sufficient (and necessary) condition
for solvability of (\ref{4.11}) in $H_1(\Omega_0)$ is the equality
$\lambda_\pm=-(f_\pm,\phi_0)_\Omega$.

It remains to get the relations (\ref{4.10}). We denote
$B^{\pm}(t)=\Omega_0\backslash S_\pm(t)$. The functions
$\widetilde\psi_\pm$ satisfy the statements of
Lemma~\ref{lm3.1}; integrating by parts the left hand sides of
the equalities
$$
-\left((\Delta+\lambda_0)\psi_\pm,\phi_0\right)_{B^\pm(t)}
=\lambda_\pm(\phi_0,\phi_0)_{B^\pm(t)},
$$
taking into account (\ref{4.8}) and the asymptotics $\phi_0(x)$
and passing to the limit as $t\to0$, we obtain the relations
(\ref{4.10}).
\end{proof}

In view of Lemmas~\ref{lm4.1} and \ref{lm3.3} the function
$\phi_2$ defined by the equality (\ref{1.8}) is a solution of the
boundary value problem (\ref{3.4}) for $ \lambda_2$ defined by the
equality (\ref{1.6}).

Thus, the coefficients $\lambda_2$, $\phi_2$, $v_0^\pm$ and
$v_1^{\pm}$ of the series (\ref{1.1})--(\ref{1.3}) satisfying the
statement of Theorem~\ref{th1.2} have been constructed.

\section{Inner expansion}

We shall seek the complete inner expansion for the eigenfunction
in the form (\ref{1.4}). Since the following construction of the
coefficients of the inner expansions is the same for both
endpoints of the slit, then, firstly, we consider only
expansions at the left endpoint, and, secondly, to avoid
cumbersome expressions we omit the superscripts ``$\pm$'' where
possible. Substituting (\ref{1.2}) and (\ref{1.4}) into
(\ref{1.1}), passing to the inner variables and writing down the
equality of the same power of $\varepsilon$, one can obtain the
following recursive system of the boundary value problems
\begin{equation}\label{5.1}
-\Delta v_n=\sum_{k=0}^{n-4}\lambda_k v_{n-k-4},\quad
\xi\in\Pi,\qquad \frac{\partial v_n}{\partial\nu}=0, \quad
\xi\in\partial\Pi.
\end{equation}

We denote by  $\zeta_1$ and $\zeta_2$ the real and imaginary
parts of the complex variable
$w=\sqrt{\xi_1-\frac{1}{4}+\mathrm{i}\xi_2}$, $\mathrm{i}$ is
the imaginary unit. In variables $\zeta=(\zeta_1,\zeta_2)$ the
boundary value problem (\ref{5.1}) becomes simpler
\begin{align}
-\Delta_\zeta v_n&=|\zeta|^2\sum_{k=0}^{n-4}\lambda_k v_{n-k-4},
\quad \zeta_2>\frac{1}{2},\label{5.2}
\\
\frac{\partial v_n}{\partial\zeta_2}&=0,\quad
\zeta_2=\frac{1}{2}.\label{5.3}
\end{align}
\begin{lemma}\label{lm5.1} For each natural $k$

a) the boundary value problem
$$
\Delta v=0,\quad \zeta_2>\frac{1}{2},\qquad \frac{\partial
v}{\partial\zeta_2}=\zeta^k_1,\quad \quad \zeta_2=\frac{1}{2}
$$
has a solution of the form
$$
v(\zeta)=\sum_{j=0}^{[k/2]}\alpha_j\mathrm{Im}\,w^{k+1-2j},
$$
where $\alpha_j$ are some explicitly calculated constants,
$\alpha_0=\frac{1}{k+1}$;

b) there exists a solution of a boundary value problem
$$
\Delta_\zeta Y_{k+1}=0,\quad \zeta_2>\frac{1}{2}, \qquad
\frac{\partial Y_{k+1}}{\partial\zeta_2}=0,\quad
\zeta_2=\frac{1}{2}
$$
that can be represented in the form
$$
Y_n(\zeta)=\mathrm{Re}\,w^n+\sum_{j=0}^{n-2} \beta_j\mathrm{Im}\,
w^{n-1-2j},
$$
where $\beta_j$ are some explicitly calculated constants. The
leading term of the asymptotics for the function
$X_n(\xi)=Y_n(\zeta_1(\xi),\zeta_2(\xi))$ as $\xi\to\infty$ has
the form $\rho^{n/2}\cos(n\theta/2)$.
\end{lemma}
\begin{proof} The validity of the item a) follows from
the equality
$$
\frac{\partial}{\partial\zeta_2}\mathrm{Im}\,w^n\Bigg|_{\zeta_2=
\frac{1}{2}}=n\sum_{j=0}^{\left[\frac{n-1}{2}\right]}C^{2j}_{n-1}
\zeta_1^{n-1-2j}\left(-\frac{1}{4}\right)^j.
$$
in the obvious way. In its turn, the item a) and an equality
$$
\frac{\partial}{\partial\zeta_2}\mathrm{Re}\,w^n\Bigg|_{\zeta_2=
\frac{1}{2}}=-\frac{n}{2}
\sum_{j=0}^{\left[\frac{n}{2}\right]-1}C^{2j+1}_{n-1}
\zeta_1^{n-2-2j}\left(-\frac{1}{4}\right)^j.
$$
yield the validity of the item b).
\end{proof}

For the sake of uniformity of notations, we set $X_0=Y_0=1$.
Note that under these notations the functions $v_k^\pm$
constructed above have the form $v_0^\pm=\phi^\pm_0X_0$,
$v_1^\pm=d_\pm g^\pm X_1+\phi^\pm_1X_0$.

\begin{lemma}\label{lm5.2} Let  $0\le n<\infty$, $0\le k\le n$, and
$\{a_k^{(n)}\}$, $\lambda_n$ be arbitrary sequences of real
numbers. Then the system of the boundary value problems
(\ref{5.2}), (\ref{5.3}) has the system of solutions represented
in the form
\begin{align}
v_n(\zeta)=&\sum_{i=0}^na_i^{(n)}Y_i(\zeta) +
\widetilde{v}_n(\zeta),\label{5.4}
\\
\widetilde{v}_n(\zeta)=&\sum_{k=1}^{\left[\frac{n}{4}\right]} \|
w\|^{4k}\left(\sum_{j= 0}^{n-4k}
\alpha_{n,k,j}\mathrm{Re}\,w^j+\sum_{j= 2}^{n-4k}
\beta_{n,k,j}\mathrm{Im}\,w^{j-1}\right)+
\\
&+\sum_{j=2}^n\beta_{n,0,j}\mathrm{Im}\,w^{j-1}\nonumber
\end{align}
where the constants $\alpha_{n,k,j}$ and $\beta_{n,k,j}$ do not
depend on $\lambda_m$ as $m> n-4k-j$ and on $a^{(m)}_s$ as
$m>n-4k-j+s$.
\end{lemma}
\begin{proof} The proof is carried out by induction. For $n\le3$
equations (\ref{5.2}) are homogeneous and, in view of the item
b) of Lemma~\ref{lm5.1}, there exist solutions of the form
(\ref{5.4}) with $\widetilde v_n\equiv0$. For $n\ge4$ equations
(\ref{5.2}) are inhomogeneous and in order to construct their
solutions one has to bear in mind that
$$
\Delta\left(|w|^{4k}\mathrm{Im}\,w^j\right)=4\mathrm{Im}\,\frac{\partial^2}
{\partial\overline {w}\partial w} \left(\overline {w}^{2k}
w^{2k+j}\right)=8k(2k+j)|w|^{4k-2}\mathrm{Im}\,w^{j}
$$
and a similar equality holds for $|w|^{4k}\mathrm{Re}\,w^{j}$. For
this reason, it easy to construct  the solution of the
inhomogeneous solution (\ref{5.2}) of the form (\ref{5.4}),
moreover, without the last sum in the representation for
$\widetilde v_n$. To eliminate the discrepancy appeared in the
boundary condition (\ref{5.3}) one should use the item b) of
Lemma~\ref{lm5.1}, what imply the appearance of the last sum in
the representation for $\widetilde v_n$. The independence of the
constants $\alpha_{n,k,j}$ and $\beta_{n,k,j}$ on $a^{(m)}_s$ and
$\lambda_m$ for the changing of their indexes in ranges mentioned
in the lemma follows from the algorithm of construction $v_n$
which has been adduced.
\end{proof}

We denote $z=\xi_1-\frac{1}{4}+\mathrm{i}\xi_2$, where
$\mathrm{i}$ is an imaginary unit. By definition, $z=w^2$, what
and Lemma~\ref{lm5.2} imply
\begin{lemma}\label{lm5.3} Let  $0\le n<\infty$, $0\le k\le n$,
and $\{a_k^{(n)}\}$, $\lambda_n$ be arbitrary sequences of real
numbers.  Then the system of the boundary value problems
(\ref{5.1}) has the system of solutions represented in the form
\begin{equation}
\begin{aligned}
v_n(\xi)=&\sum_{i=0}^na_i^{(n)}X_i(\xi)+\widetilde{v}_n(\xi),
\\
\widetilde{v}_n(\xi)=&\sum_{k=1}^{\left[\frac{n}{4}\right]} \|
z\|^{2k}\left(\sum_{j= 0}^{n-4k}
\alpha_{n,k,j}\mathrm{Re}\,z^{j/2}+\sum_{j= 2}^{n-4k}
\beta_{n,k,j}\mathrm{Im}\,z^{(j-1)/2}\right)+
\\
&+\sum_{j=2}^n\beta_{n,0,j}\mathrm{Im}\,z^{(j-1)/2},
\end{aligned}\label{5.5}
\end{equation}
where the constants $\alpha_{n,k,j}$ and $\beta_{n,k,j}$ do not
depend on $\lambda_m$ as $m> n-4k-j$ and on $a^{(m)}_s$ as
$m>n-4k-j+s$.
\end{lemma}

From Lemma~\ref{lm5.3} it follows that for any
$\lambda_\varepsilon$, having power asymptotics with arbitrary
coefficients $\lambda_j$, the series
$$
v_\pm(x;\varepsilon)=\sum_{n=0}^\infty\varepsilon^nv_n
\left(\frac{x_\pm}{\left(g^\pm\varepsilon\right)^2}\right),
$$
whose coefficients satisfy the statements of Lemma~\ref{lm5.3},
are asymptotic solutions of (\ref{1.1}) near the endpoints of the
slit. In order to define the true values of the constants
$\lambda_j$ and $a^{(m)}_s$ one must consider the outer expansions
in the vicinity of the endpoints of the slit and to compare
(match) it with the inner expansion constructed.

\section{Outer expansion}

We seek the asymptotics for the eigenfunction outside a
neighbourhood of the endpoints of the slit (the outer expansion)
in the form of the series (\ref{1.3}). Observe that the
asymptotics expansion (\ref{1.3}) (similarly to the asymptotics
expansions (\ref{1.4})) corresponds to the eigenfunction
$\phi_\varepsilon$ with ``lax'' normalization
$\left\|\phi_\varepsilon\right\|_{0,\Omega}=1+o(1)$ as
$\varepsilon\to 0$.

The boundary value problems for the coefficients of the series
(\ref{1.3}) are obtained in the standard way. We substitute the
series (\ref{1.2}) and (\ref{1.3}) into (\ref{1.1}) and then we
write down the equalities of the same power of $\varepsilon$ and
formally pass to limit as $\varepsilon\to0$. As a result, we get
the following recursive system of the boundary value problems
\begin{equation}
\begin{aligned}
-\Delta\phi_n=&\sum^n_{j=0}\lambda_j\phi_{n-j},\quad x\in\Omega_0,
\qquad\frac{\partial\phi_n}{\partial\nu}=0,\quad x\in\Gamma,
\\
\frac{\partial}{\partial x_2}\phi_n(x_1,\pm0)=
&-\sum_{j=1}^n\frac{1}{j!}\left(g_\pm(x_1)\right)^j
\frac{\partial^{j+1}\phi_{n-j}}{\partial x_2^{j+1}}(x_1,\pm0)
\\
&+ g'_\pm(x_1)\sum_{j=0}^{n-1}\frac{1}{j!}
\left(g_\pm(x_1)\right)^j\frac{\partial^{j+1}\phi_{n-j-1}}{\partial
x_2^j\partial x_1}(x_1,\pm0),\quad x_1\in\omega_0.
\end{aligned}\label{6.1}
\end{equation}

The aim of this section is to study the solvability of the
problems of the form (\ref{6.1}) in a class of singular
solutions.

Let $j$ be any half-integer, $H_j(x)$  be homogeneous functions of
order $j$ belonging to $C^\infty(\mathbb{R}^2\backslash l_{+})$,
where $l_{+}$ is the semiaxis $x_2=0$, $x_1\ge0$. We denote by
$\widetilde{\mathcal{H}}_m$ the set of the series of the form
\begin{equation}
H(x)=\sum_{j=-m}^\infty H_{j/2}(x).\label{6.2}
\end{equation}
We call the terms of negative order the singular part of the
series.

Similarly, we denote by $\widetilde{\mathcal{A}}_m$ the set of the
series of the form
$$
h(x)=\sum_{j=-m}^\infty \alpha_jx_1^{j/2}.
$$

\begin{definition*} The scalar sequence
$$
\left\{b_j=\frac{1}{2\pi r^{j/2}}\int\limits_0^{2\pi}
H_{j/2}(x)\cos\left(\frac{j\theta}{2}\right)\,d\theta\right\}_{j=0}^{\infty}
$$
is called a harmonic sequence of the series (\ref{6.2}).
\end{definition*}
\begin{lemma}\label{lm6.1}
Let the series $H\in\widetilde{\mathcal{H}}_0$ have the zero
harmonic series and be a formal asymptotic solution as $r\to0$
of the boundary value problem
$$
(\Delta+\lambda_0)H = 0, \quad x\in\mathbb{R}^2\backslash l_+,
\qquad \frac{\partial}{\partial x_2}H(x_1,\pm0)=0,\quad x_1>0.
$$
Then $H=0$.
\end{lemma}

In proving this lemma one should bear in mind that the terms of
the series belonging to $\widetilde{\mathcal{H}}_0$ have the form
$r^{j/2}\Phi_j(\theta)$, $j\ge0$. After the substitution in the
equation we get an ordinary differential equation for $\Phi_j$.
The explicit form of the solutions of these equations, the
boundary conditions as $\theta=0$, $\theta=2\pi$ and the fact that
$b_j$ equals zero yield the statement of the lemma.
\begin{corollary*}
Let $F\in\widetilde{\mathcal H}_m$, $h_\pm\in\widetilde{\mathcal
A}_m$, and let the series $H^{(1)},\,H^{(2)}\in\widetilde{\mathcal
H}_m$ be formal asymptotic solutions as $r\to0$ of a boundary
value problem
$$
(\Delta+\lambda_0)H = F, \quad x\in{\mathbb R}^2\backslash l_+,
\qquad\frac{\partial}{\partial x_2} H(x_1,\pm0)=h_\pm(x_1),\quad
x_1>0,
$$
and let these series have the same harmonic sequences and
$H^{(1)}-H^{(2)}\in\widetilde{\mathcal H}_0$. Then
$H^{(1)}=H^{(2)}$.
\end{corollary*}

We denote by ${\mathcal H}_m$ the subset of the functions in
$C^\infty(\overline\Omega_0\backslash\{O_+;O_-\})$ whose
asymptotics at the points $O_\pm$ belong to the class
$\widetilde{\mathcal H}_m$ (with respect to the coordinate systems
$x_+$ and $x_-$). Similarly, let ${\mathcal A}_m$ be the subset of
functions in $C^\infty\left(\omega\right)$ with asymptotic
behaviour at the endpoints of the slit described by functions in
the class $\widetilde{\mathcal A}_m$.

We consider a boundary value problem
\begin{equation}\label{6.3}
(\Delta+\lambda_0)u=F+\lambda\phi_0,\quad x\in\Omega_0,\qquad
\frac{\partial}{\partial x_2}u(x_1,\pm0)=h_\pm(x_1), \quad x_1\in
\omega_0,
\end{equation}
where $F\in{\mathcal H}_m$, $h_\pm\in{\mathcal A}_m$, $m\in\mathbb
N$.
\begin{lemma}\label{lm6.2}
Suppose that two classes of series
$H_\pm(x;\{b_j\}_{j=0}^\infty,\lambda)$ belong to
$\widetilde{\mathcal H}_m$ for each value of the parameters $b_j$
and $\lambda$ and have the following properties

a) the series $H_\pm(x_\pm;\{b_j\}_{j=0}^\infty,\lambda)$ are
asymptotics solutions of the problem (\ref{6.3}) for $x_\pm\to0$;

b)  $\{b_j\}_{j=0}^\infty$ is a harmonic sequence of the series
$H_\pm(x;\{b_j\}_{j=0}^\infty,\lambda)$;

c) $H_\pm(x;\{\widetilde b_j\}_{j=0}^\infty,\lambda)-
H_\pm(x;\{\widehat
b_j\}_{j=0}^\infty,\lambda)\in\widetilde{\mathcal H}_0$ for each
sequences $\{\widetilde b_j\}_{j=0}^\infty$ and $\{\widehat
b_j\}_{j=0}^\infty$.

Then there exist numbers $\lambda$ and
$\{b^\pm_j\}_{j=0}^\infty$ and a function $u\in{\mathcal H}_m$
such that $u$ is a solution of the boundary value problem
(\ref{6.3}) and it has asymptotics coinciding with
$H_\pm(x_\pm;\{b_j^\pm\}_{j=0}^\infty,\lambda)$ as $x_\pm\to0$.
\end{lemma}
\begin{proof}
This statement is proved by arguments similar to those used in the
proof of Lemma~\ref{lm4.1}. We seek the solution of the boundary
value problem (\ref{6.3}) in the form
\begin{equation}\label{6.4}
u(x)=u_N(x)=\chi(r_+)H_N^+(x_+)+\chi(r_-)H_N^-(x_-)+U_N(x),
\end{equation}
where  $H^\pm_N$ are partial sums (to the powers
$r^N_\pm$,inclusive) of the series $\widehat H_\pm(x)=
H_\pm(x;\{b_j=0\}_{j=0}^\infty,0)$. Substituting (\ref{6.4}) into
(\ref{6.3}), we deduce a boundary value problem for $u_N$:
\begin{equation}\label{6.5}
(\Delta+\lambda_0)U_N=F_N+\lambda\phi_0,\quad x\in\Omega_0,\qquad
\frac{\partial}{\partial x_2}U_N(x_1,\pm0)=h_N^\pm(x_1), \quad
x_1\in\omega_0.
\end{equation}
Here the function $F_N\in C^{N-2}(\overline\Omega_0)$ (where,
recall, the cut $\gamma_0$ is interpreted as double-sided) has
zeroes of order $r_\pm^{N-2}$ at the endpoints of the slit, and
the functions $h_N^+,\,h_N^-\in
C^{N-1}\left(\overline\omega_0\right)$ have zeroes of order
$x_1^{N-1}$ and $(1-x_1)^{N-1}$ at the corresponding endpoints of
the interval $\omega_0$. Therefore, there exists a constant
$\lambda=\lambda(N)$ for which the boundary value problem
(\ref{6.5}) is solvable in the functional class $H_1(\Omega_0)$.
On the other hand, substituting $u_{N_1}-u_{N_2}\in H_1(\Omega_0)$
into (\ref{6.3}), one can easy see that, firstly, $\lambda$ does
not depend on $N$, and, secondly, the functions $u_N$ are the same
for different values of $N$ up to the term equal to the
eigenfunction. Due to the arbitrariness in choosing $N$, the
results of \cite{Kn} and Lemma~\ref{lm3.1} we conclude that there
exists a solution $u\in{\mathcal H}_m$ of the boundary value
problem (\ref{6.3}) (for some constant $\lambda$), having
asymptotics at the endpoints of the slit of the form
$$
u(x)=\widetilde H_\pm(x_\pm),\quad x_\pm\to0,
$$
where $\widetilde H_\pm$ differs from $\widehat H_\pm$ by a term
in $\widetilde{\mathcal H}_0$.

Let $\{b_j^\pm\}$ be a harmonic sequence of the series $\widetilde
H_\pm$. Form the Corollary to Lemma~\ref{lm6.1} it follows that
the series $\widetilde H_\pm(x_\pm)$ and
$H_\pm(x_\pm;\{b_j^\pm\}_{j=0}^\infty,\lambda)$ are the same.
\end{proof}

\section{Matching the expansions}

We introduce re-expansions operators ${\mathcal M}^\pm$ on the
formal series of the type
$$
V(\xi;\varepsilon)=\sum_{n=0}^\infty\varepsilon^nV_n(\xi)
$$
by the following standard procedure. Coefficients of the series
$V$ are replaced by their asymptotics at infinity and then we pass
to the variables $x_\pm=(\varepsilon{g^\pm})^2\xi$. The formal
double series obtained is called the value of ${\mathcal
M}^\pm(V(\xi;\varepsilon))$.

For the sake of brevity we use the notations $\widetilde{\mathcal
H}_{-1}=\widetilde{\mathcal H}_0$, $\mathcal H_{-1}={\mathcal
H}_0$. From Lemma~\ref{lm5.3} and the definition of the
re-expansion operators it follows
\begin{lemma}\label{lm7.1} Let all asumptions of Lemma~\ref{lm5.3} hold
and the coefficients of the series
$$
v(\xi;\varepsilon)=\sum_{n=0}^\infty\varepsilon^nv_n(\xi)
$$
satisfy the statement of Lemma~\ref{lm5.3}. Then the
representation
$$ {\mathcal
M}^\pm\left(v(\xi;\varepsilon)\right)=\sum^\infty_{j=0}\varepsilon^j\Phi_j^\pm(x_\pm).
$$
is true. The series $\Phi_n^\pm\in\widetilde{\mathcal H}_{n-1}$
are formal asymptotic solutions (as $x_\pm\to0$) of the recursive
system of boundary value problems (\ref{6.1}), where the functions
$\phi_j$ in the right hand sides of the equations and boundary
conditions are replaced by $\Phi_j^\pm$.

A harmonic sequence
$\{\overset{\pm}{b}_i\!\!\big.^{(n)}\}_{i=0}^\infty$ of a series
$\Phi_n^\pm$ has the form
$\overset{\pm}{b}_i\!\!\big.^{(n)}=a^{(n+i)}_i
\left(g^\pm\right)^{-i}$.

The series $\Phi_n^\pm$ do not depend on $a^{(m+i)}_i$ and
$\lambda_m$ for $m>n$, and their singular parts also do not depend
on $a^{(n+i)}_i$ and $\lambda_n$.
\end{lemma}

We denote by  $v^\pm(\xi;\varepsilon)$ the series (\ref{1.4}).

\begin{theorem}\label{th7.1} There exist series (\ref{1.2})--(\ref{1.4})
having the following properties

a) $\phi_n\in{\mathcal H}_{n-1}$ are the solutions of (\ref{6.1});

b) $v_n^\pm$ are the solutions of the boundary value problems
(\ref{5.1}) and they can be represented in the form (\ref{5.5})
(with the indexes ''$\pm$'' added);

c) ${\mathcal M}^\pm\left(v^\pm(\xi;\varepsilon)\right)=
\sum^\infty_{j=0}\varepsilon^j\phi_j(x)$ as $x_\pm\to0$;

d) the coefficients $\lambda_1$, $\lambda_2$, $\phi_1$, $\phi_2$,
$v_0^\pm$ and $v_1^\pm$ satisfy the statements of
theorem~\ref{th1.2}.
\end{theorem}
\begin{proof} Let $v_n^\pm(\xi)$ satisfy the statements of
Lemma~\ref{5.3} with constants $\lambda_j$,
$a^{(m)}_s=\overset{\pm}{a}_s\!\!\big.^{(m)}$ undefined yet. Then,
in view of Lemma~\ref{lm7.1},
$$
{\mathcal
M}^\pm\left(v^\pm(\xi;\varepsilon)\right)=\sum^\infty_{j=0}\varepsilon^j\Phi_j^\pm(x_\pm).
$$
For $n\le2$ we denote by $\widetilde\Phi_n^\pm(x_\pm)$ the
asymptotics as $x_\pm\to0$ of the solutions $\phi_n$ of the
boundary value problems (\ref{3.2})--(\ref{3.4}) those are
defined above, and we denote by
$\{\overset{\pm}{b}_i\!\!\big.^{(n)}\}_{i=0}^\infty$ their
harmonic sequences. In construction of the coefficients
$v_j^\pm$ we set $\overset{\pm}{a}_i\!\!\big.^{(n+i)}
=\overset{\pm}{b}_i\!\!\big.^{(n)}\left(g^\pm\right)^i$, and we
define $\lambda_n$ in accordance with the formulae (\ref{1.5})
and (\ref{1.6}). Lemma~\ref{lm7.1} and Corollary to
Lemma~\ref{lm6.1} imply that $\Phi_0^\pm=\widetilde\Phi_0^\pm$,
$\Phi_1^\pm=\widetilde\Phi_1^\pm$. Note that having defined
$\overset{\pm\quad}{a^{(n+i)}_i}$ we determine $v_n^\pm$ and
three leading harmonics for other $v^\pm_j$. From the structure
(\ref{5.5}) of the functions $v_i^\pm$ (more precisely, from the
form of $v_1^\pm$) it follows that
$\Phi_2^\pm-\widetilde\Phi_2^\pm\in\widetilde{\mathcal H}_0$.
Thus, by Corollary to Lemma~\ref{lm6.1} we obtain
$\Phi_2^\pm=\widetilde\Phi_2^\pm$. It should be stressed that
having defined $v_2^\pm$, due to Lemma~\ref{lm7.1} we determined
singular parts of the series $\Phi_3^\pm\in\widetilde{\mathcal
H}_2$.

In next step due to Lemma~\ref{lm6.2} by singular parts of the
asymptotics series $\Phi_3^\pm$ we define the function
$\phi_3\in\mathcal H_2$ which is a solution of (\ref{6.1}) for
some value of $\lambda_3$ and whose asymptotics as $x_\pm\to0$
coincide with $\Phi_3^\pm$ for some
$\overset{\pm}{a}_i\!\!\big.^{(3+i)}=\overset{\pm}{b}_i\!\!\big.^{(3)}
\left(g^\pm\right)^i$, etc.
\end{proof}

The only fact following from the items a) and b) of
Theorem~\ref{th7.1} is the series (\ref{1.3}) and (\ref{1.2}) are
asymptotic solutions of the problem (\ref{1.1}) for
$r_+>\varepsilon$, $r_->\varepsilon$, and the series (\ref{1.4})
and (\ref{1.2}) are asymptotic solutions of the problem
(\ref{1.1}) for $r_\pm<2\varepsilon$. The key condition of
matching is determined by the item c) of the theorem proved.

\section{Justification of the asymptotics}

We use the notations $\lambda_{\varepsilon,N}$,
$\phi_{\varepsilon,N}(x)$ and
$v^\pm_{\varepsilon,N}\left(x_\pm\left(\varepsilon
g^\pm\right)^{-2}\right)$ for the partial sums of the series
(\ref{1.2})--(\ref{1.4}). Further, we set
\begin{align*}
\Phi_{\varepsilon,N}(x)=&\left(1-\chi(r_+\varepsilon^{-1})\right)
\left(1-\chi(r_-\varepsilon^{-1})\right)\phi_{\varepsilon,N}(x)+\\
&+\chi(r_+\varepsilon^{-1})v^+_{\varepsilon,N}\left(x_+\left(\varepsilon
g^+\right)^{-2}\right)+
\chi(r_-\varepsilon^{-1})v^-_{\varepsilon,N}\left(x_-\left(\varepsilon
g^-\right)^{-2}\right).
\end{align*}
\begin{lemma}\label{lm8.1} Suppose that the series
(\ref{1.2})--(\ref{1.4}) satisfy the statements of
Theorem~\ref{th7.1}. Then the function $\Phi_{\varepsilon,N}(x)$
is a solution of a boundary value problem
\begin{equation}\label{8.1}
\begin{aligned}
-&(\Delta+\lambda_{\varepsilon,N})\Phi_{\varepsilon,N} =
f_{\varepsilon,N}, \quad x\in\Omega_\varepsilon,
\\
&\frac{\partial}{\partial\nu}\Phi_{\varepsilon,N}=0,\quad
x\in\Gamma,\qquad\frac{\partial}{\partial\nu}\Phi_{\varepsilon,N}=
h_{\varepsilon,N},\quad x\in \gamma_\varepsilon,
\end{aligned}
\end{equation}
where $\| f_{\varepsilon,N}\|_{0,\Omega}\le C_N\varepsilon^{M}$,
$h_{\varepsilon,N}=O(\varepsilon^M)$ in the norm of
$C^1(\gamma_\varepsilon)$, and $M\to\infty$ as $N\to\infty$.
\end{lemma}
\begin{proof} The statement of the lemma being a standard
implication of the items a)--c) of Theorem~\ref{th7.1} (see, for
instance, \cite{Il2}), we give a brief proof. Substituting
$\Phi_{\varepsilon,N}$ in the left hand sides of (\ref{8.1}), we
get that a homogeneous boundary condition on $\Gamma$ holds and
the functions $f_{\varepsilon,N}$, $h_{\varepsilon,N}$ can be
represented in the form
$$
f_{\varepsilon,N}=\sum_{i=1}f_{\varepsilon,N}^{(i)},\qquad
h_{\varepsilon,N}=\sum_{i=1}h_{\varepsilon,N}^{(i)},
$$
where
\begin{align*}
f_{\varepsilon,N}^{(1)}=&-\left(1-\chi(r_+\varepsilon^{-1})\right)
\left(1-\chi(r_-\varepsilon^{-1})\right)
\sum_{n=1}^N\sum_{k=N+1-n}^N\varepsilon^{k+n}\lambda_n\phi_k,
\\
f_{\varepsilon,N}^{(2)}=&-\sum_{n=1}^N\sum_{k=N-3-n}^N\varepsilon^{k+n}\lambda_n\left(
\chi(r_-\varepsilon^{-1})v_k^-+\chi(r_+\varepsilon^{-1})v_k^+\right),
\\
f_{\varepsilon,N}^{(3)}=&\sum_{i=1}^3\frac{\partial}{\partial
x_i}\left(\phi_{\varepsilon,N}(x)-v^-_{\varepsilon,N}\left(x_-\left(\varepsilon
g^-\right)^{-2}\right)\right)\frac{\partial}{\partial
x_i}\chi(r_-\varepsilon^{-1})
\\
&+\sum_{i=1}^3\frac{\partial}{\partial
x_i}\left(\phi_{\varepsilon,N}(x)-v^+_{\varepsilon,N}\left(x_+\left(\varepsilon
g^+\right)^{-2}\right)\right)\frac{\partial}{\partial
x_i}\chi(r_+\varepsilon^{-1})
\\
&
+\left(\phi_{\varepsilon,N}(x)-v^-_{\varepsilon,N}\left(x_-\left(\varepsilon
g^-\right)^{-2}\right)\right)\Delta \chi(r_-\varepsilon^{-1})
\\
&
+\left(\phi_{\varepsilon,N}(x)-v^+_{\varepsilon,N}\left(x_+\left(\varepsilon
g^+\right)^{-2}\right)\right)\Delta \chi(r_+\varepsilon^{-1}),
\\
h_{\varepsilon,N}^{(1)}=& \left(1-\chi(r_+\varepsilon^{-1})\right)
\left(1-\chi(r_-\varepsilon^{-1})\right)\frac{\partial}{\partial\nu}\phi_{\varepsilon,N}(x),\\
h_{\varepsilon,N}^{(2)}=&
\left(\phi_{\varepsilon,N}(x)-v^-_{\varepsilon,N}\left(x_-\left(\varepsilon
g^-\right)^{-2}\right)\right)\frac{\partial}{\partial
\nu}\chi(r_-\varepsilon^{-1})
\\
&+\left(\phi_{\varepsilon,N}(x)-v^+_{\varepsilon,N}\left(x_+\left(\varepsilon
g^+\right)^{-2}\right)\right)\frac{\partial}{\partial
\nu}\chi(r_+\varepsilon^{-1}).
\end{align*}
In view of the item a) of Theorem~\ref{th7.1} the functions
$f_{\varepsilon,N}^{(1)}$ and $h_{\varepsilon,N}^{(1)}$ have the
norms of order $O(\varepsilon^{M_1})$ in $H_0(\Omega)$ and in
$C^1(\gamma_\varepsilon)$, respectively, and $M_1\to\infty$ as
$N\to\infty$. Similarly, by the item b) we deduce that
$f_{\varepsilon,N}^{(2)}$ has a norm of order
$O(\varepsilon^{M_2})$ in $C^1(\gamma_\varepsilon)$, and
$M_2\to\infty$ as $N\to\infty$. Finally, the item c) of
Theorem~\ref{th7.1} implies that norms of
$f_{\varepsilon,N}^{(3)}$ and $h_{\varepsilon,N}^{(2)}$ have order
$\varepsilon^{M_3}$ in $H_0(\Omega)$ and
$C^1(\gamma_\varepsilon)$, respectively, and $M_3\to\infty$ as
$N\to\infty$. These facts completes the proof for
$M=\min\{M_1;M_2;M_3\}$.
\end{proof}

The boundary condition on $\gamma_\varepsilon$ in (\ref{8.1})
being inhomogeneous, we can not apply Lemma~\ref{lm2.6} directly
to (\ref{8.1}) in order to justify the asymptotics. For this
reason, beforehand we shall prove two auxiliary statements.
\begin{lemma}\label{lm8.2} Let $u\in
C^2(\Omega_\varepsilon)\cap C^1 (\overline\Omega_\varepsilon)$,
$\frac{\partial}{\partial\nu}u>0$ on $\gamma_\varepsilon$, $\Delta
u<0$ in $\Omega_\varepsilon$ and $u\ge0$ on $\Gamma$. Then $u\ge0$
in $\overline\Omega_\varepsilon$.
\end{lemma}
\begin{proof} Since $\frac{\partial}{\partial\nu}u>0$ on
$\gamma_\varepsilon$, we conclude that the minimum of $u$ lies
outside $\gamma_\varepsilon$, and, as $\Delta u<0$ in
$\Omega_\varepsilon$, then it can not lie in
$\Omega_\varepsilon$. Hence, the minimum lies in $\Gamma$,
where, by conditions, $u\ge0$.
\end{proof}

\begin{lemma}\label{lm8.3} Let $U\in C^2(\Omega_\varepsilon)\cap C^1
(\overline\Omega_\varepsilon)$,
$$
\max\limits_\Gamma|U|+
\max\limits_{\gamma_\varepsilon}\left|\frac{\partial
U}{\partial\nu}\right|+\sup\limits_{\Omega_\varepsilon}|\Delta
U|=m.
$$
Then $|U|<Cm\varepsilon^{-1}$, where $C$ is some constant
independent on $U$.
\end{lemma}
\begin{proof} Recall that near the left endpoint of the slit
(for $0\le x_1<t_0$, where $t_0$ is some positive constant) the
equation of $\gamma_\varepsilon$ has the form $\varepsilon^2
x_1=\left(g^-\right)^{-2}x_2^2$. Similarly, near the right
endpoint of the slit (for $0\le 1-x_1<t_0$) its equation  reads as
follows $\varepsilon^2(1- x_1)=\left(g^+\right)^{-2}x_2^2$. For
$t_0\le x_1\le 1-t_0$ the equation of $\gamma_\varepsilon$ has the
form $x_2=\varepsilon g_\pm(x_1)$. Note that
$d=\min\limits_{t_0\le x_1\le 1-t_0}|g_\pm(x_1)|>0$ and $\pm
g_\pm(x_1)>0$ as $t_0\le x_1\le 1-t_0$. We set
$V(x)=\left(x_1-\frac{1}{2}\right)^2+\alpha^2x_2^2$, where
$\alpha>0$ is some constant.

Then for $t_0\le x_1\le1-t_0$ one can check
\begin{equation}\label{8.2}
\begin{aligned}
{}&\frac{\partial V}{\partial \nu}\Bigg|_{x_2=\varepsilon
g_\pm}=\mp\left(1+\varepsilon^2\left(g'_\pm(x_1)\right)^2\right)^{-1/2}
\left(\varepsilon
g_\pm'(x_1)\left(1-2x_1\right)+2\alpha^2\varepsilon
g_\pm(x_1)\right)
\\
{}&=\mp2\varepsilon\left(1+\varepsilon^2\left(g'_\pm(x_1)\right)^2\right)^{-1/2}
\left(\alpha^2 g_\pm(x_1)-\left(x_1-\frac{1}{2}\right)
g_\pm'(x_1)\right)<-\varepsilon\alpha^2 d
\end{aligned}
\end{equation}
for $\varepsilon$ sufficiently small and $\alpha$ chosen
appropriately. In its turn, for  $0\le x_1<t_0$ we have
\begin{equation}\label{8.3}
\begin{aligned}
\frac{\partial V}{\partial \nu}\Bigg|_{x_1=
\left(\frac{x_2}{\varepsilon g^-}\right)^2} =&
\left(\varepsilon^4+4\left(g^-\right)^{-4}x_2^2\right)^{-1/2}\left(\varepsilon^2\frac{\partial
V}{\partial x_1}-2\frac{x_2}{(g^-)^2}\frac{\partial V}{\partial
x_2}\right)\Bigg|_{x_1= \left(\frac{x_2}{\varepsilon
g^-}\right)^2}
\\
=&
\varepsilon^{-1}\left(\varepsilon^4+4\left(g^-\right)^{-2}x_1\right)^{-1/2}
\left(\varepsilon^2\left(2x_1-1\right)-4\varepsilon^2\alpha^2
x_1\right)
\\
=&-\varepsilon\left(\varepsilon^4+4\left(g^-\right)^{-2}x_1\right)^{-1/2}
(1-2x_1+4\alpha^2 x_1)<\varepsilon C_-<0.
\end{aligned}
\end{equation}
%
Similarly, for $1-t_0 <x_1\le 1$ we deduce that
\begin{equation}\label{8.4}
\frac {\partial V}{\partial \nu}\Bigg|_{1-x_1=
\left(\frac{x_2}{\varepsilon g^+}\right)^2}<\varepsilon C_+<0.
\end{equation}

We set $C=\max|C_\pm|$, $W(x)=R-V(x)$, where
$R>1+\max\limits_{\overline\Omega}|V(x)|$. Then the estimates
(\ref{8.2})--(\ref{8.4}) imply that the functions
$Cm\varepsilon^{-1}W\pm U$ satisfy the hypothesis of
Lemma~\ref{lm8.2}. In its turn, Lemma~\ref{lm8.2} yields the
correctness of the statement being proved.
\end{proof}

\emph{Proof of Theorem~\ref{th1.1}} Let $\chi_\Gamma\in
C^\infty_0(\Omega)$ be a function equal to one outside some
neighbourhood of $\Gamma$, $U_{\varepsilon,N}\in
C^\infty(\overline{\Omega}_\varepsilon)$ be a harmonic function
satisfying boundary condition
$$
U_{\varepsilon,N}=0,\quad x\in\Gamma,\qquad
\frac{\partial}{\partial\nu}U_{\varepsilon,N}=h_{\varepsilon,N},
\quad x\in\gamma_\varepsilon,
$$
$\widetilde \Phi_{\varepsilon,N}=\chi_\Gamma U_{\varepsilon,N}$.
Then from Lemmas~\ref{lm8.1},~\ref{lm8.3} and well-known a priori
estimates it follows that
\begin{equation}\label{8.5}
\begin{aligned}
{}&\frac{\partial}{\partial\nu}\widetilde\Phi_{\varepsilon,N}=0,\quad
x\in\Gamma, \qquad
\frac{\partial}{\partial\nu}\widetilde\Phi_{\varepsilon,N}=h_{\varepsilon,N},\quad
x\in \gamma_\varepsilon,
\\
{}&\| \widetilde \Phi_{\varepsilon,N}\|_{1,\Omega_\varepsilon}+
\|\Delta \widetilde
\Phi_{\varepsilon,N}\|_{0,\Omega_\varepsilon} \le
C_N\varepsilon^{M-1}.
\end{aligned}
\end{equation}
We set $\widehat\Phi_{\varepsilon,N}=
\Phi_{\varepsilon,N}-\widetilde\Phi_{\varepsilon,N}$. By
Lemma~\ref{lm8.1} and relations (\ref{8.5}), we obtain that the
function $\widehat\Phi_{\varepsilon,N}$ is a solution of a
boundary value problem
\begin{equation}\label{8.6}
-\Delta\widehat\Phi_{\varepsilon,N} =
\lambda_{\varepsilon,N}\widehat\Phi_{\varepsilon,N} +
F_{\varepsilon,N}, \quad x\in\Omega_\varepsilon,
\qquad\frac{\partial}{\partial\nu}\widehat
\Phi_{\varepsilon,N}=0,\quad x\in\partial \Omega_\varepsilon,
\end{equation}
where
\begin{equation}\label{8.7}
\| F_{\varepsilon,N}\|_{0,\Omega}\le C_N\varepsilon^{M}.
\end{equation}
Since $\Phi_{\varepsilon,N}\to\phi_0$ в $H_0(\Omega_0)$, then from
(\ref{8.5}) it follows that
\begin{equation}\label{8.8}
\widehat\Phi_{\varepsilon,N}\to\phi_0\quad\text{в
$H_0(\Omega_0)$}.
\end{equation}
In view of (\ref{8.6})--(\ref{8.8}), Lemma~\ref{lm2.4}, the
estimate (\ref{2.15}), and the arbitrariness in choosing $N$ we
conclude that the series (\ref{1.2}) constructed coincides with
the asymptotics of the eigenvalue $\lambda_\varepsilon$. In
their turn, the estimates (\ref{2.16}) and (\ref{8.8}) imply
that
\begin{equation}\label{8.9}
\begin{aligned}
\widehat\Phi_{\varepsilon,N}(x)&=\alpha_{\varepsilon,N}\phi_\varepsilon(x)+
\widetilde\phi_{\varepsilon,N}(x),
\\
\|\widetilde\phi_{\varepsilon,N}
\|_{0,\Omega}&=O(\varepsilon^M),\quad |\alpha_{\varepsilon,N}|
\to1,\qquad\varepsilon\to0 .
\end{aligned}
\end{equation}
Finally, due to (\ref{8.5}), (\ref{8.9}) and the arbitrariness in
choosing $N$ we deduce that the asymptotic series (\ref{1.3}),
(\ref{1.4}) are the asymptotics of the eigenfunction
$\phi_\varepsilon$ associated with the eigenvalue converging to
$\lambda_0$.\quad\endproof

\section{Concluding remarks}

The asymptotics of eigenvalues in the case of Dirichlet boundary
condition on $\gamma_\varepsilon$ was considered in \cite{GI}.
There it was shown that if the boundary of  the slit lies on
square parabolas near the endpoints, then the asymptotics of the
eigenelements have the power character (\ref{1.2})--(\ref{1.4}).
However, for the case the equation of the endpoints of the slit
have the form of square parabolas only ``in principal'' it was
shown in \cite{GI} that the asymptotics of $\lambda_\varepsilon$
and $\phi_\varepsilon$ contain also powers of $\ln\varepsilon$.
It can be shown that this effect takes place in the case of
Neumann boundary condition. It also can established that for the
Robin boundary condition on $\gamma_\varepsilon$ the powers of
logarithms appear in asymptotic expansions even in the case when
the boundary of slit lie on parabolas near the endpoints.

\end{document}